\documentclass[twocolumn,english,superscriptaddress,10pt,aps,pra,showpacs]{revtex4-1}

\usepackage[T1]{fontenc}
\usepackage[utf8]{luainputenc}
\usepackage{graphicx}
\usepackage{amsmath}
\usepackage{wasysym}
\usepackage{mathtools}

\usepackage{subfigure}

\usepackage{parskip}
\usepackage[colorlinks,
            linkcolor=black,
            anchorcolor=blue,
            citecolor=blue,
            urlcolor=blue,
            bookmarks=false
            ]{hyperref}

\makeatletter

\@ifundefined{textcolor}{}
{%
 \definecolor{BLACK}{gray}{0}
 \definecolor{WHITE}{gray}{1}
 \definecolor{RED}{rgb}{1,0,0}
 \definecolor{GREEN}{rgb}{0,1,0}
 \definecolor{BLUE}{rgb}{0,0,1}
 \definecolor{CYAN}{cmyk}{1,0,0,0}
 \definecolor{MAGENTA}{cmyk}{0,1,0,0}
 \definecolor{YELLOW}{cmyk}{0,0,1,0}
}

\makeatother

\usepackage{babel}

\begin{document}

\title{Modulation of single-photon-level wave packets with two-component electromagnetically induced transparency}

\author{Sheng-Jun Yang}
\affiliation{Hefei National Laboratory for Physical Sciences at Microscale and Department of Modern Physics, University of Science and Technology of China, Hefei, Anhui 230026, China}
\affiliation{CAS Center for Excellence and Synergetic Innovation Center in Quantum Information and Quantum Physics, University of Science and Technology of China, Hefei, Anhui 230026, China}

\author{Xiao-Hui Bao}
\affiliation{Hefei National Laboratory for Physical Sciences at Microscale and Department of Modern Physics, University of Science and Technology of China, Hefei, Anhui 230026, China}
\affiliation{CAS Center for Excellence and Synergetic Innovation Center in Quantum Information and Quantum Physics, University of Science and Technology of China, Hefei, Anhui 230026, China}

\author{Jian-Wei Pan}
\affiliation{Hefei National Laboratory for Physical Sciences at Microscale and Department of Modern Physics, University of Science and Technology of China, Hefei, Anhui 230026, China}
\affiliation{CAS Center for Excellence and Synergetic Innovation Center in Quantum Information and Quantum Physics, University of Science and Technology of China, Hefei, Anhui 230026, China}

\date{\today}

\begin{abstract}
Coherent manipulation of single-photon wave packets is essentially important for optical quantum communication and quantum information processing. In this paper, we realize controllable splitting and modulation of single-photon-level pulses by using a tripod-type atomic medium. The adoption of two control beams enable us to store one signal pulse into superposition of two distinct atomic collective excitations. By controlling the time delay between the two control pulses, we observe splitting of a stored wave packet into two temporally-distinct modes. By controlling the frequency detuning of the control beams, we observe both temporal and frequency-domain interference of the retrieval signal pulses, which provides a method for pulse modulation and multi-splitting of the signal photons.
\end{abstract}


\maketitle

Techniques for tailoring the dispersion of materials, especially the electromagnetically induced transparency (EIT) method, enable us to control light propagation better than before and inspire lots of applications in quantum optics and atomic physics~\cite{Lukin2003, Fleischhauer2005}. The ordinary EIT structure is a three-level atomic system interacting with two laser fields, named the signal and control light. By dynamically decreasing/increasing the control light power, coherent state of the signal photons could be transferred into/from the atoms, described as the atom-photon dark-state polaritons~\cite{Fleischhauer2000, Fleischhauer2002}. For now, EIT has become an important method for photonic quantum memory\,---\,an essential component of quantum information science~\cite{Sangouard2011, Pan2012}. Furthermore, coherent control of the spinor polaritons would lead to numbers of useful quantum state engineering.

Particularly, a tripod-type four-level EIT scheme, which offers more freedom and possibilities for photon storage and manipulation, has drawn significant attentions in the last few years~\cite{Paspalakis2002, Petrosyan2004, Rebic2004, Mazets2005, Wu2005, Raczynski2006, Raczynski2007, Han2008, Karpa2008, Li2008, Aghamalyan2010, Wang2011, Sowik2012}. In the tripod-EIT configuration, three ground atomic levels are coupled with an excited level by two signal fields and one control field~\cite{Karpa2008} or one signal field and two control fields. As there exist a pair of orthogonal dark-state polaritons, whose coherence depends on the amplitude and phase of the control fields, beam splitting of a signal pulse could be achieved based on this tripod-type photon storage~\cite{Raczynski2007, Aghamalyan2010}. In comparison with other beam-splitting methods~\cite{Wang2004, Xiao2008, Hosseini2009}, the tripod-EIT scheme highlights the simplicity and high degree of controllability. For instance, in a former experiment that using rapid coherence transport in a wall-coated atomic vapor cell~\cite{Xiao2008}, ratio of pulse splitting is rather difficult to control effectively. Besides, the tripod-EIT scheme offers a novel way for the modulation of single-photon wave packets~\cite{Bao2011}, which would greatly benefit quantum information processing of the photonic qubits.

By now, in the previous tripod-EIT experiments~\cite{Karpa2008, Wang2011}, constructive and destructive interference of the two-component dark-state polaritons has been observed, e.g., experimental demonstration of dual-channel memory and compensation of Larmor precession by adjusting the relative phase between the two reading beams~\cite{Wang2011}. Yet the signal pulses in all these experiments are strong classical light, and no well-separated temporal splitting and modulation of single-photon-level pulses has been achieved, which is important for the photonic time-bin qubit generation and manipulation. In this paper, by using a tripod-EIT medium, we experimentally map a signal pulse in the single-photon regime into two atomic spinwave excitations, and demonstrate a time-bin beam splitter for the retrieved signal photons by switching on the two control fields at different time instants. Furthermore by changing the detuning of the control fields, we also observe the temporal and frequency-domain oscillation of the signal photons during the retrieval process. This technique provides us a method to modulate the signal pulse shape, and thus could improve quantum interference between the photonic qubits and interactions with atoms. Such controllable temporal modulation and multi-mode splitting of the photonic wave packets in our system may find applications in quantum information processing and communication, especially for the robust time-bin quantum communication architecture~\cite{Brendel1999, Tittel2000, Inoue2002}.

\begin{figure}[tbh]
  \centering
      \includegraphics[width=\columnwidth]{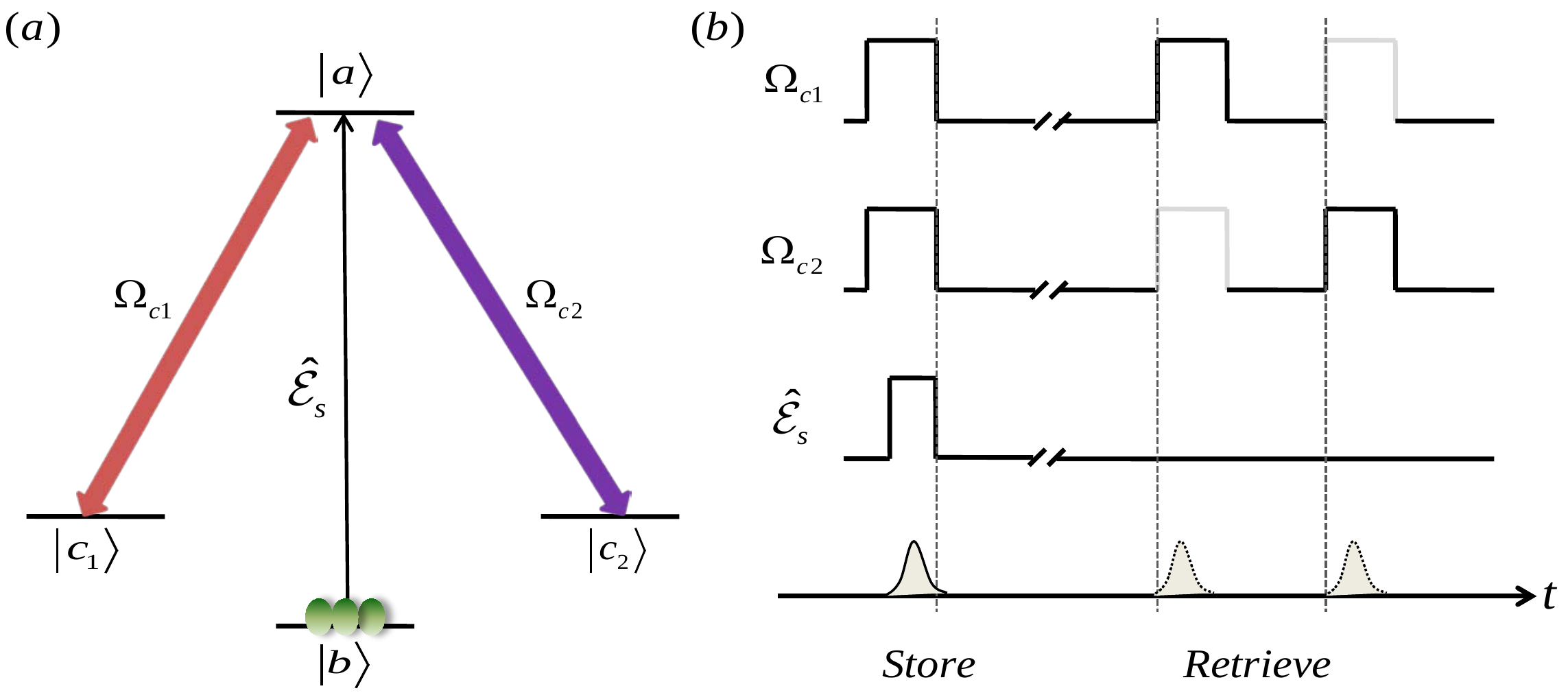}
      \caption{(color online). (a) Scheme of the tripod-EIT energy levels. The atom ensemble is coupled with one quantum signal field $\hat{\mathcal{E}}_{s}$ and two strong control fields with the Rabi frequencies $\Omega_{c1}$ and $\Omega_{c2}$, where the atoms are initially in the ground state $\left|b\right>$; (b) Time sequence of the photonic time-bin beam splitter. First, a signal photon pulse is coherently stored in the atomic spinwave excitations while $\Omega_{c1}$ and $\Omega_{c2}$ are adiabatically switched off at the same time. Then the stored signal photon is time-entangled by successively switching on these two control fields.}
      \label{fig:TripodEnergy}
\end{figure}

As shown in Fig.~\ref{fig:TripodEnergy}(a), we label the tripod-type atomic structure with one excited state $\left|a\right>$ and three independent ground states $\left|b\right>$, $\left|c_{1}\right>$ and $\left|c_{2}\right>$. All the atoms are initially located in the ground state $\left|b\right>$. A signal quantum field $\hat{\mathcal{E}}_{s}$ is resonant with the atomic transition $\left|b\right>\!\to\!\left|a\right>$, satisfying the single-photon detuning $\delta_s\!=\!0\,\rm{MHz}$. Another two strong control fields, with the Rabi frequencies $\Omega_{c1}$ and $\Omega_{c2}$, are related to the transitions $\left|c_{1}\right>\!\to\!\left|a\right>$ and $\left|c_{2}\right>\!\to\!\left|a\right>$ respectively. Assuming the single-photon detuning of the control fields $\delta_{c1}\!=\!\delta_{c2}=0\,\rm{MHz}$, we get the steady eigenstate of the atom-photon polaritons as following~\cite{Raczynski2007},
\begin{equation}
\label{eq:steadypolariton} \Psi\!=\!\cos\theta\hat{\mathcal{E}}_s\!-\!\sqrt{N}\sin\theta(\sin{\phi}\rm{e}^{-i\varphi_{c1}^0}\hat{\sigma}_{bc1}\!+\!\cos{\phi}\rm{e}^{-i\varphi_{c2}^0}\hat{\sigma}_{bc2}),
\end{equation}
where $\tan{\theta}\!=\!\sqrt{N}g/\left|\Omega\right|$, $g\!=\!\sqrt{\hbar\omega/2\epsilon_0V}$, $N$ is the total atom number, and $\Omega$ is Rabi frequency of the control fields, $\left|\Omega\right|^2\!=\!\left|\Omega_{c1}\right|^2\!+\!\left|\Omega_{c2}\right|^2$. The atomic component of the polaritons has two spinwave excitations $\hat{\sigma}_{bc1}$ and $\hat{\sigma}_{bc2}$, where $\tan{\phi}\!=\!\Omega_{c1}/\Omega_{c2}$, and $\varphi_{ci}^0$ ($i\!=\!1,2$) is the phase of the control field $\Omega_{ci}$ that transferred onto the atoms. During the EIT storage process, the signal photon is mapped into these two spinwave excitations which behave as an atomic qubit and are completely determined by the control fields. As shown in Fig.~\ref{fig:TripodEnergy}(b), a time-bin superposition state can be generated by successively retrieving the two atomic excitations $\hat{\sigma}_{bc1}$ and $\hat{\sigma}_{bc2}$ at a storage time $t_1$ and $t_2$,
\begin{equation}
\label{eq:timebinEntangled}
\left|\Phi\right>_{t_1,t_2}\!=\!\sin{\phi}\left|1_{t_1}0_{t_2}\right>\!+\!\cos{\phi}\rm{e}^{i\delta\varphi_{c}}\left|0_{t_1}1_{t_2}\right>,
\end{equation}
where the relative phase, $\delta\varphi_c\!=\!\varphi_{c1}^{t_1}\!-\!\varphi_{c2}^{t_2}\!+\!\varphi_{c1}^{0}\!-\!\varphi_{c2}^{0}$, is determined by the two control fields.

\begin{figure}[tbh]
  \centering
      \includegraphics[width=\columnwidth]{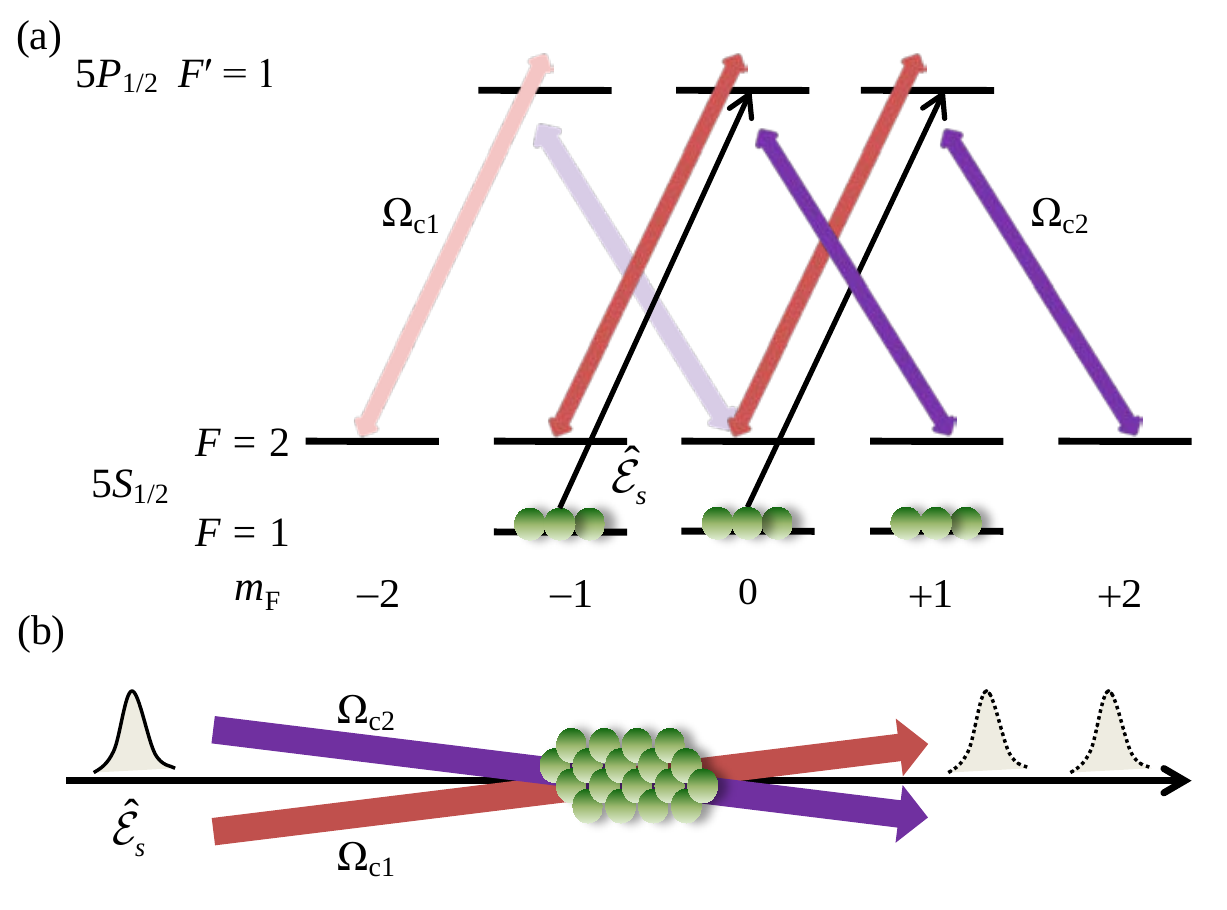}
      \caption{(color online). (a) The $^{87}\rm{Rb}$ atomic levels of the tripod-EIT configuration; (b) Schematic of the experiment setup.}
      \label{fig:ExpSetup}
\end{figure}

In the experiment, the tripod-type system is a $^{87}\rm{Rb}$ cold atomic ensemble loaded by the magnetic optical trapping (MOT). The stray magnetic field is compensated with three orthogonal Helmholtz coils. All the atoms are initially prepared in the degenerate ground state $\left|5S_{1/2},F\!=\!1\right>$, with an optical depth about 1.3(1) corresponding to the $D1$-line transition $\left|F\!=\!1\right>\!\to\!\left|F'\!=\!1\right>$. As shown in Fig.~\ref{fig:ExpSetup}(a), a right circularly polarized ($\sigma^+$) signal pulse, with an envelope of $\hat{\mathcal{E}}_{s}$, is resonant of the transition $\left|F\!=\!1\right>\!\to\!\left|F'\!=\!1\right>$; and there is less than one signal photon per pulse on average. Two strong control fields, that of left and right circularly polarization ($\sigma^{\mp}$) respectively, are resonant or near resonant of the $D1$-line $\left|F\!=\!2\right>\!\to\!\left|F'\!=\!1\right>$, connecting the signal pulse with two sub-Zeeman levels of the ground state $\left|F\!=\!2\right>$. The signal and control fields, from two laser diodes, are phase-locked with each other. Clearly, there exist two independent groups of the tripod-EIT structure, connecting the Raman coherence $\left|F\!=\!1,m_F\!=\!-1\right>\!\leftrightarrow\!\left|2,-1\right>\&\left|2,+1\right>$ and $\left|1,0\right>\!\leftrightarrow\!\left|2,0\right>\&\left|2,+2\right>$ respectively. The three light beams are arranged in the same horizontal plane, as shown in Fig.~\ref{fig:ExpSetup}(b). The signal pulse is sent through the atoms, with a beam waist diameter $\sim\!600\,\rm{\mu m}$, and detected by a single photon detector. The two control fields are forward propagating with a small angle of the signal field, and beam diameters of them are $\sim\!2\,\rm{mm}$, to cover the whole signal beam. Time resolution of the measurement system is 2\,ns, thus we could obtain pulse shape of the signal photons.

\begin{figure}[tbh]
  \centering
  \subfigure{
    \label{fig:TripodSplitWaveform} 
    \includegraphics[width=\columnwidth]{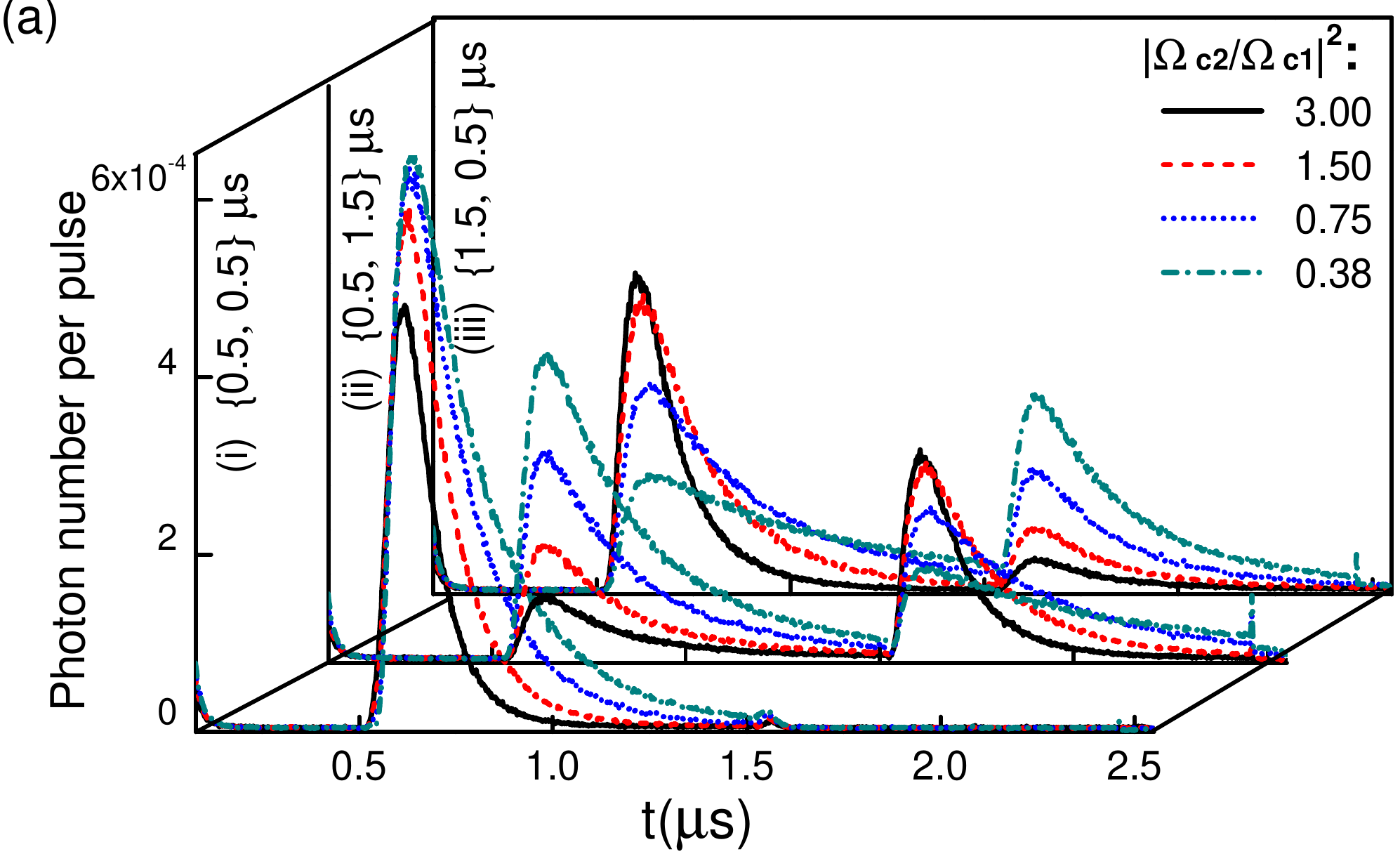}}
  \hspace{0.0in}\\
  \subfigure{
   \label{fig:TripodSplitRate}
    \includegraphics[width=\columnwidth]{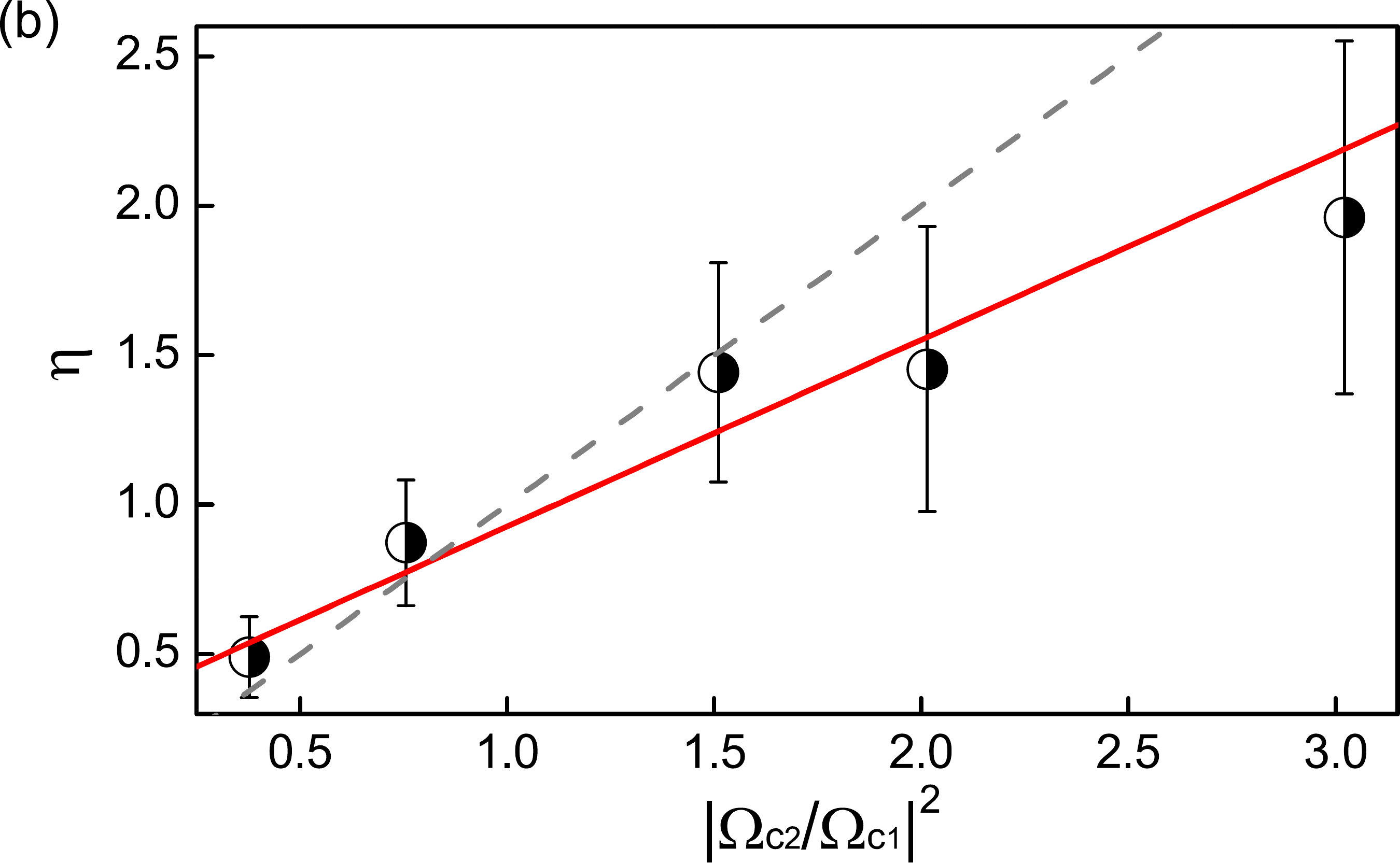}}
  \caption{(color online.) (a) Pulse shape of the retrieved signal photon splitting, where the $y$-axis is the statistic retrieved photon count per pulse. From the front to back of the three subgraphs, time sequence of the control fields during the retrieval process are, (i) both $\Omega_{c1}$ and $\Omega_{c2}$ on at the storage time $0.5\,\rm{\mu s}$, (ii) $\Omega_{c1}$ on at $0.5\,\rm{\mu s}$ and $\Omega_{c2}$ at $1.5\,\rm{\mu s}$, and (iii) vice versa. In each subgraph, laser power of the control field $\Omega_{c2}$ is varying, while $\Omega_{c1}$ keeps constant of $100\,\rm{\mu W}$; (b) Splitting proportion of the retrieved signal pulses, corresponding to the setting (iii) in (a). The $x$-axis is the ratio of the laser power between the two control fields; and the $y$-axis is the photon splitting proportion $\eta\!=\!n_{c2}/n_{c1}$, where $n_{c1}$ ($n_{c2}$) is the total photon counts retrieved by the control field $\Omega_{c1}$ ($\Omega_{c2}$). The red solid line is linear fitting of the experimental results, and the slop of the grey dashed line equals one.}
  \label{fig:PhotonSplitting} 
\end{figure}

First, we verify time-bin splitting of the signal photons in memory. All the three laser beams are resonant with their relative atomic transitions. Time sequence of the experiment is shown in Fig.~\ref{fig:TripodEnergy}(b), with the same general strategy in~\cite{Wang2011}. At the begining, the control fields $\Omega_{c1}$ and $\Omega_{c2}$ are simultaneously switched off to map a signal pulse into the atomic spinwave excitations. After a storage time $0.5\,\rm{\mu s}$, one control pulse $\Omega_{c1}$ ($\Omega_{c2}$) is switched on again, lasting $1\,\rm{\mu s}$ to retrieve the corresponding spinwave excitations. Subsequently, the second control pulse $\Omega_{c2}$ ($\Omega_{c1}$) is on for another $1\,\rm{\mu s}$ to retrieve the rest of the atomic excitations. This means that the storage time of the first retrieval component is $0.5\,\rm{\mu s}$ and the second $1.5\,\rm{\mu s}$.

Based on theoretical discussion above, beam splitting is related to the control field power. The weight of the spinwave components $\hat{\sigma}_{bc1}$ and $\hat{\sigma}_{bc2}$ in the polariton $\Psi$, depends on the Rabi frequencies $\Omega_{c1}$ and $\Omega_{c2}$. In Fig.~\ref{fig:PhotonSplitting}(a), we show the temporal splitting of the retrieved signal pulses. There are four settings of the control laser power, that changing the control pulse power $\Omega_{c2}$ while keeping $\Omega_{c1}$ constant at $100\,\rm{\mu W}$. As atoms in our system are equally distributed in the ground Zeeman states, if we open the field $\Omega_{c1}$ first, the stored signal photons by different control fields could be retrieved independently in time sequence, as shown in Fig.~\ref{fig:PhotonSplitting}(a)ii. But if we open $\Omega_{c2}$ first, part of the signal photons stored by $\Omega_{c2}$ would be destroyed. This could be avoided if we optically pump all the atoms into a single Zeeman state. From envelope of the retrieved wave packets shown in Fig.~\ref{fig:PhotonSplitting}(a), proportion of the signal pulse retrieved by $\Omega_{c1}$ is dropping while increasing the laser power $\Omega_{c2}$. Compared with former demonstration of dual-channel selective retrieval~\cite{Wang2011}, we have successfully retrieved the two stored photon components in sequence. Furthermore in order to see clearly the beam splitting effect, we calculate the retrieval proportion $\eta$ for different settings of the control laser power in Fig.~\ref{fig:PhotonSplitting}(b). The grey dashed line corresponds to the ideal situation that the storage efficiencies of the signal photons induced by the two control fields are the same. Because group velocity of the signal pulse $v_g\propto\left|\Omega\right|^2$, the EIT storage efficiency decreases while the control light power is too strong. Thus for large laser power of the control field $\Omega_{c2}$, the proportion $\eta$ is below the value of $\left|\Omega_{c2}/\Omega_{c1}\right|^2$.

By tuning the relative Rabi frequencies of the two control pulses, we have successfully split a signal pulse into two parts that propagating in the same direction at a controllable storage time. As the signal pulse is simultaneously stored in the two spinwave excitations which are phase preserving with each other~\cite{Wang2011}, the two splitting photon components are in coherence. Thus to retrieve the two components at the same time, the control fields need to meet the relative phase of the two spinwave excitations; otherwise, the spinwave excitations would remain in the atoms. This would provide us a method to modulate pulse shape of the stored signal photons.

As we know theoretically for the tripod-EIT configuration, the other orthogonal spinor polariton is
\begin{equation}
\Psi^\perp\!=\!\sqrt{N}(\cos{\phi}\rm{e}^{-i\varphi_{c2}^0}\hat{\sigma}_{bc1}\!-\!\sin{\phi}\rm{e}^{-i\varphi_{c1}^0}\hat{\sigma}_{bc2}),
\end{equation}
which doesn't interact with the signal pulse $\hat{\mathcal{E}}_s$ and $\left<\Psi^\perp|\Psi\right>\!=\!0$. If laser power of the two control fields keeps constant during the storage and retrieval processes, but frequency detuning of them are different, that $\delta_{c1}-\delta_{c2}\!=\!2\delta_c$, $\delta_{c1}+\delta_{c2}\!=\!2\delta_s$, evolution of the two spinwave excitations satisfies
\begin{equation}
\label{eq:polaritonevolution}
-i\frac{\partial\Psi}{\partial t}\!=\!\left(\delta_s-\delta_c\cos2\phi\right)\Psi\!+\!\delta_c\sin2\phi\Psi^\perp.
\end{equation}
When we change the relative phase of the two control fields, the polaritons $\Psi$ and $\Psi^\perp$ would be transformed, similar to the optical waveplate rotation. Only the component that in phase with the control fields could be released out of the atomic medium. Assuming $\delta_s\!=\!0\,\rm{MHz}$, $\delta_c\!\neq\!0\,\rm{MHz}$ and $\left|\Omega_{c1}\right|\!=\!\left|\Omega_{c2}\right|$, the two atomic excitations alternately evolve into each other, and would create a temporal oscillation of the retrieved signal pulse with an oscillating period of $2\pi/\delta_c$.

\begin{figure}[h]
  \centering
  \subfigure{
    \label{fig:BeatingWaveform} 
    \includegraphics[width=\columnwidth]{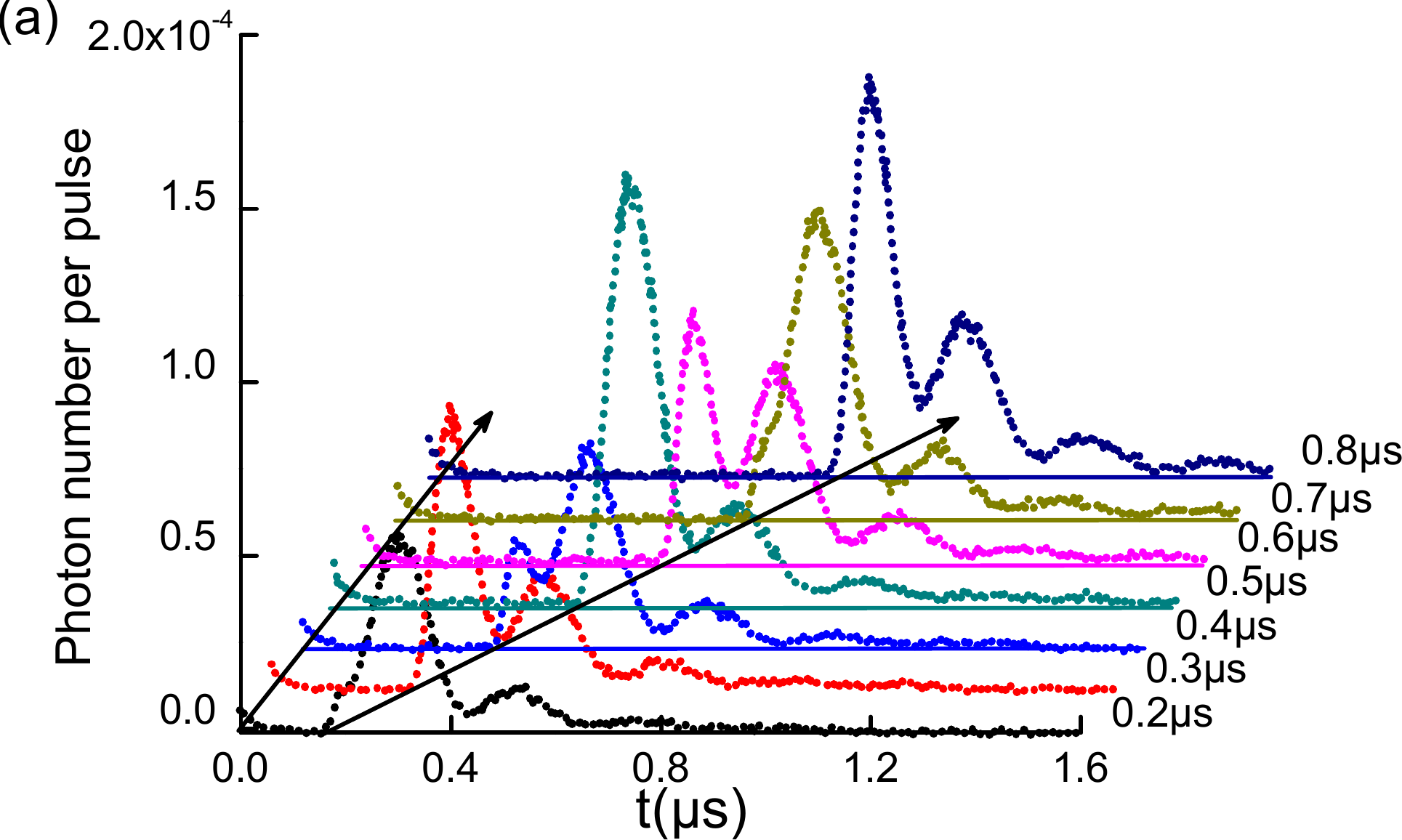}}
\\
\hspace{0.0in}
  \subfigure{
    \label{fig:BeatingFreq} 
    \includegraphics[width=\columnwidth]{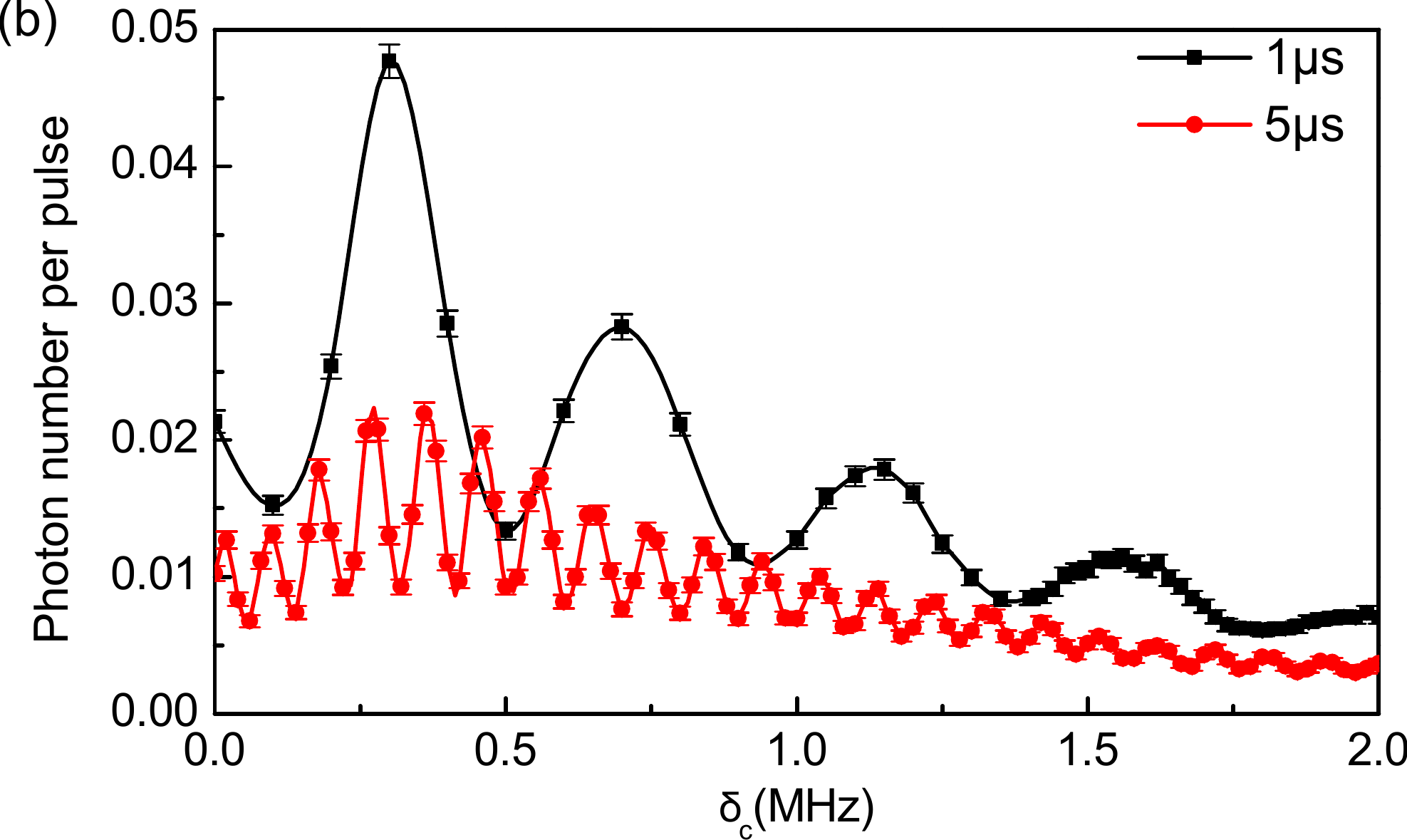}}
  \caption{(color online). (a) Pulse shape oscillation of the retrieved signal photons for various storage time $\left(0.2-0.8\right)\,\rm{\mu s}$, while setting the control field detuning $\delta_c\!=\!\delta_{c1}\!=\!-\delta_{c2}\!=\!2.0\,\rm{MHz}$. The period of the temporal oscillations is about $0.25\,\rm{\mu s}$; (b) Statistics of the retrieved signal photon counts per pulse while changing the control fields' detuning $\delta_c$ at the storage time of $1\,\rm{\mu s}$ (black square) and $5\,\rm{\mu s}$ (red circle). The $y$-axis is the sum of the signal photons within the initial $200\,\rm{ns}$ time interval of the retrieved pulse.}
  \label{fig:PolaritonInterference} 
\end{figure}

To identify such temporal modulation of the signal pulses, in the following experiment we set the two control fields in opposite frequency detuning, that $\delta_{c1}\!=\!-\delta_{c2}\!=\!\delta_c$. As $\{\delta_{c1},\delta_{c2}\}\!\ne\!0\,\rm{MHz}$, the storage efficiency would become smaller than that of resonance. But when the detuning is small, the oscillating period is large compared with the retrieved pulse width and no obvious pulse shaping is observable in such a situation. Considering these two factors, we choose the detuning $\delta_{c}$ of $2\,\rm{MHz}$ and both of the control laser power about $100\,\rm{\mu W}$. The control fields are simultaneously on during the storage and retrieval processes. And we observe the polariton interference in the two processes. The temporal oscillation of the retrieved signal pulse is shown in Fig.~\ref{fig:PolaritonInterference}(a) for different storage time $0.2\!-\!0.8\,\rm{\mu s}$. The signal pulse exhibits series of maxima and minima intensity oscillation, and the beating period is about $250\,\rm{ns}$. Because the stationary atomic excitations and the control fields would accumulate a phase difference $\delta_c T$ during a storage time $T$, pulse shape of the retrieved signal photons are the same at the storage time $\tau\,\rm{\mu s}$ and $\left(\tau\!+\!0.25\rm{n}\right)\,\rm{\mu s}$ (n=1, 2, 3...). This shows that the stored phase information and coherence is preserved in the atomic medium. By fine adjustment of the control fields' amplitude and phase, the beating signals would be more obvious, and the destructive points could reach zero. Such waveform modulation may allow us to retrieve a sequence of Gaussian-like wave pulses with equal pulse width, more than just two pulse splitting discussed above.

For certain storage time $T$, the phase shift $\delta_c T$ varies with the control field detuning $\delta_c$. Thus at the begining of the retrieval process, the starting in-phase component of the spinwave excitations periodically oscillates while sweeping the frequency detuning $\delta_c$, and the temporal oscillation would start with a different pulse amplitude. In Fig.~\ref{fig:PolaritonInterference}(b), we sum the photon counts over the initial $200\,\rm{ns}$ interval of the retrieval signal pulse after a storage time $1\,\rm{\mu s}$ and $5\,\rm{\mu s}$, while sweeping the control field detuning $\delta_c$ from $0\,\rm{MHz}$ to $2\,\rm{MHz}$. The photon counts are oscillating with the control field detuning, with a frequency oscillating period $1/2T$. As shown in Fig.~\ref{fig:PolaritonInterference}(b), the maximum point of the retrieved photon counts is not under the resonant condition $\delta_c\!=\!0\,\rm{MHz}$. This is because the different AOM (acousto-optic modulator) switching time of the two control light results in an additional phase shift $\delta\phi_c\!\sim\!0.78\pi$ between the two spinwave components. For long coherence lifetime of the EIT memory, we could obtain much higher resolution of the frequency difference $\delta_c$, which may be useful in precise frequency measurement and metrology.

In conclusion, we have experimentally studied the propagation and storage of single-photon-level pulses in a tripod-type EIT medium. We achieved time-bin splitting of a signal pulse that is stored in two collective spinwave excitations. The splitting ratio and temporal separation are highly controllable. By changing frequency detuning of the two control fields, we have further observed both the temporal and frequency-domain interference for the retrieved signal pulses. Such waveform modulation hasn't been done in other tripod-type experiments, and is rather feasible to generate shaped photonic pulses. This technique would be useful in efficient quantum interference between photons and their interactions with atoms. By improvement of the memory efficiency~\cite{Chen2013} and the coherence lifetime~\cite{Schnorrberger2009, Dudin2010}, a better performance of the experimental results would be achievable. This work paves a way for the interface with other type of single-photon sources like quantum dots~\cite{Rakher2013} and narrowband SPDC~\cite{Bao2008} etc. Furthermore, with similar methods it is also possible to realize a pulse combiner of the time-bin qubits. We believe that manipulation of narrowband single-photon temporal modes in the tripod-EIT system would have potential applications in quantum information processing and time-bin based quantum communication~\cite{Brendel1999, Tittel2000, Inoue2002}.

\textit{Note added.}$-$After completing this work we became aware of a related experiment with a double-tripod spinor system by Lee \emph{et al.}~\cite{Lee2014}.

\section*{Acknowledgement}
This work was supported by the National Natural Science Foundation of China, National Fundamental Research Program of China (under Grant No. 2011CB921300), and the Chinese Academy of Sciences. X.-H.B. acknowledge support from the Youth Qianren Program.

\bibliography{Myrefs}
\bibliographystyle{apsrev4-1}

\end{document}